\documentclass[aps,showpacs,preprintnumbers,amsmath, amssymb]{revtex4}

\usepackage{float}
\usepackage{graphics,epsfig}
\usepackage{graphicx}
\usepackage{dcolumn}
\usepackage{bm}

\begin{document}
\baselineskip=0.8 cm

\title{{\bf Analytical studies on the hoop conjecture in charged curved spacetimes}}
\author{Yan Peng$^{1}$\footnote{yanpengphy@163.com}}
\affiliation{\\$^{1}$ School of Mathematical Sciences, Qufu Normal University, Qufu, Shandong 273165, China}

\vspace*{0.2cm}
\begin{abstract}
\baselineskip=0.6 cm
\begin{center}
{\bf Abstract}
\end{center}

Recently, with numerical methods,
Hod clarified the validity of Thorne hoop conjecture
for spatially regular static charged fluid spheres,
which were considered as counterexamples against the hoop conjecture.
In this work, 
we provide an analytical proof on Thorne hoop
conjecture in the spatially regular static
charged fluid sphere spacetimes.

\end{abstract}

\maketitle
\newpage
\vspace*{0.2cm}

\section{Introduction}

One famous conjecture in general relativity is the
Thorne hoop conjecture, which states that horizons appear when and only
when a mass $\mathcal{M}$ gets compacted into a region whose circumference C in every direction is
$C\leqslant 4\pi \mathcal{M}$ \cite{hc1,hc2}. This upper bound can be saturated in
the case of Schwarzschild black hole with the horizon radius $r_{0}=2\mathcal{M}$.
If generically true, such conjecture would signify that black holes form if matter/energy
is enclosed in a small enough region.
At present, there are a lot of works addressing the hoop
conjecture, see \cite{hc3}-\cite{ahc9} and references therein.

Intriguingly, a few counterexamples against
hoop conjecture were also presented \cite{hc19,hc20}.
In particular, Ref. \cite{hc19} constructed the horizonless charged fluid
sphere configurations with uniform charge densities.
For $\frac{M}{r_{0}}=0.65$ and $\frac{Q^2}{r_{0}^2}=0.39$,
the horizonless charged sphere satisfies a relation  $\frac{C(r_{0})}{4\pi M}\backsimeq 0.769<1$,
where $M$ is the total mass of the spacetime, Q is the
sphere charge and $r_{0}$ is the sphere radius.
According to the relation $\frac{C(r_{0})}{4\pi M}<1$
in the horizonless spcetime,
the author claimed that Thorne hoop conjecture
can be violated in horizonless charged fluid sphere spacetimes \cite{hc19}.

However, as stated by Hod, it is physically more appropriate to interpreted
the mass term $\mathcal{M}$ in Thorne hoop conjecture as the gravitational mass $M(r_{0})$ contained
within the radius $r_{0}$ and not as the total mass $M$ of the
entire curved spacetime \cite{hc21}. In fact, there is electric energy outside the charged sphere.
For the same parameters $\frac{M}{r_{0}}=0.65$ and $\frac{Q^2}{r_{0}^2}=0.39$ as \cite{hc19},
Hod reexamined the validity of hoop conjecture for charged fluid spheres and numerically obtained the
relation $\frac{C(r_{0})}{4\pi M(r_{0})}\backsimeq 1.099>1$,
which is in fact in accordance with the hoop conjecture in charged spacetime \cite{hc21}.
Along this line, it is still meaningful to analytically
examine Thorne hoop conjecture for spatial regular charged fluid spheres
with generic parameters.

The rest of the paper is organized as follows.
We shall introduce the gravity model of spatial regular static charged fluid spheres.
We provide an analytical proof on Thorne hoop conjecture
for horizonless charged fluid spheres with generic parameters.
Finally, we will briefly summarize our results.

\section{Validity of the hoop conjecture for charged fluid spheres}

It was widely believed that Thorne hoop conjecture is a
fundamental property of classical general relativity.
A lot of works indeed support the hoop conjecture \cite{hc3}-\cite{hc18}.
However, counterexamples against hoop conjecture
were also presented in \cite{hc19,hc20}.
In particular, Ref. \cite{hc19} constructed a gravity model of
horizonless static charged fluid spheres.
And the charged fluid sphere reads \cite{metric1,metric2,metric3,metric4}
\begin{eqnarray}\label{AdSBH}
ds^{2}&=&e^{\nu}dt^{2}-e^{\lambda}dr^{2}-r^{2}(d\theta^2+sin^{2}\theta d\phi^{2}).
\end{eqnarray}
The interior metric solution with uniform charge density was obtained in \cite{hc19} as
\begin{eqnarray}\label{AdSBH}
e^{-\lambda}=[1-(1-f_{0})\upsilon^2]^2,
\end{eqnarray}
\begin{eqnarray}\label{AdSBH}
e^{\nu}=\frac{1}{4f_{0}}[(1-\varepsilon-f_{0})\upsilon^2+(1-3\varepsilon+f_{0}+\frac{2Q^2}{r_{0}^2})]^2e^{\lambda/2}.
\end{eqnarray}
We define $\upsilon=\frac{r}{r_{0}}\leqslant 1$, $\varepsilon=\frac{M}{r_{0}}$ and
$f_{0}=\sqrt{1-\frac{2M}{r_{0}}+\frac{Q^2}{r_{0}^2}}$,
where $M$ is the total mass of the spacetime, Q is the
sphere charge and $r_{0}$ is the sphere radius.

In the exterior region $r\geqslant r_{0}$, the background is the
Reissner-Nordsr$\ddot{o}$m solution given by \cite{RN1,RN2,RN3,RN4,RN5,RN6}
\begin{eqnarray}\label{AdSBH}
e^{\nu}=e^{-\lambda}=1-\frac{2M}{r}+\frac{Q^2}{r^2}.
\end{eqnarray}
At the radius $r_{0}$, interior metric (2-3) and exterior metric (4) coincide with each other.

Thorne hoop conjecture states that horizons appear when and only
when a mass $\mathcal{M}$ gets compacted into a region whose circumference C in every direction is
$C\leqslant 4\pi \mathcal{M}$ \cite{hc1,hc2}.
The author in \cite{hc19} claimed that Thorne hoop conjecture was violated
by a relation
\begin{eqnarray}\label{AdSBH}
\frac{C(r_{0})}{4\pi M}\backsimeq 0.769<1
\end{eqnarray}
with $\frac{M}{r_{0}}=0.65$ and $\frac{Q^2}{r_{0}^2}=0.39$.
However, in the charged background, as stated in \cite{hc21},
it is physically more appropriate to interpreted
the mass term $\mathcal{M}$ in Thorne hoop conjecture as the gravitational mass $M(r_{0})$ contained
within the radius $r_{0}$ and not as the total mass $M$ of the
entire curved spacetime. With the same parameters $\frac{M}{r_{0}}=0.65$ and $\frac{Q^2}{r_{0}^2}=0.39$,
Hod reexamined the model and numerically obtained the relation for horizonless spheres as
\begin{eqnarray}\label{AdSBH}
\frac{C(r_{0})}{4\pi M(r_{0})}\backsimeq 1.099>1,
\end{eqnarray}
which is in accordance with the hoop conjecture in charged spacetime \cite{hc21}.

In the following, we provide an analytical proof on
Thorne hoop conjecture for generic parameters.
According to the weak energy condition (WEC),
the energy density is nonnegative \cite{metric1,metric2}.
In this work, we can take a more general condition
that the energy density is real (negative or nonnegative).
The energy density of interior region is \cite{hc19}
\begin{eqnarray}\label{AdSBH}
\rho_{m}=\frac{1-f_{0}}{8\pi r_{0}^2}\{6-\upsilon^2[5(1-f_{0})+\frac{Q^2}{r_{0}^2(1-f_{0})}]\}.
\end{eqnarray}
It can be transformed into
\begin{eqnarray}\label{AdSBH}
\rho_{m}=-\frac{5\upsilon^2}{8\pi r_{0}^2}(1-f_{0})^2+\frac{3}{4\pi r_{0}^2}(1-f_{0})+\frac{Q^2\upsilon^2}{8\pi r_{0}^4}.
\end{eqnarray}
We put $1-f_{0}$ in the general form
\begin{eqnarray}\label{AdSBH}
1-f_{0}=a+bi,
\end{eqnarray}
where $a,b$ are real numbers.

Putting (9) into (8), we arrive at the relation
\begin{eqnarray}\label{AdSBH}
\rho_{m}=(-\frac{5\upsilon^2a^2}{8\pi r_{0}^2}-\frac{5\upsilon^2b^2}{8\pi r_{0}^2}+\frac{3a}{4\pi r_{0}^2}+\frac{Q^2\upsilon^2}{8\pi r_{0}^4})
+(-\frac{5\upsilon^2a}{4\pi r_{0}^2}+\frac{3}{4\pi r_{0}^2})bi.
\end{eqnarray}

At the center $\upsilon=\frac{r}{r_{0}}=0$, (10) yields
\begin{eqnarray}\label{AdSBH}
\rho_{m}(0)=\frac{3a}{4\pi r_{0}^2}
+\frac{3b}{4\pi r_{0}^2}i.
\end{eqnarray}

Since $\rho_{m}(0)$ is real, there is
\begin{eqnarray}\label{AdSBH}
b=0.
\end{eqnarray}
That is to say $f_{0}=\sqrt{1-\frac{2M}{r_{0}}+\frac{Q^2}{r_{0}^2}}$ is real according to (9), which yields
\begin{eqnarray}\label{AdSBH}
\frac{2M}{r_{0}}-\frac{Q^2}{r_{0}^2}\leqslant 1.
\end{eqnarray}
In the case of $\frac{2M}{r_{0}}-\frac{Q^2}{r_{0}^2}= 1$,
there is $1-\frac{2M}{r_{0}}+\frac{Q^2}{r_{0}^2}=0$ and
$r_{0}$ is horizon according to (4). Since we
study horizonless sphere, the spacetime should satisfies
\begin{eqnarray}\label{AdSBH}
\frac{2M}{r_{0}}-\frac{Q^2}{r_{0}^2}< 1.
\end{eqnarray}

There is electric energy outside the charged sphere.
In order to calculate $M(r_{0})$,
we should subtract the exterior electric energy $E(r>r_{0})$ from
the total energy M. Outside the sphere, one has the Maxwell
field energy density $\rho(r)=\frac{Q^2}{8\pi r^4}$.
The mass $E(r>r_{0})$ of the Maxwell field above the radius $r_{0}$ is given by
\begin{eqnarray}\label{AdSBH}
E(r>r_{0})=\int_{r_{0}}^{+\infty}4\pi r'^{2}\rho(r')dr'=\frac{Q^2}{2r_{0}}.
\end{eqnarray}
So the mass $M(r_{0})$ within the radius $r_{0}$ is \cite{hc21,mr1,mr2}
\begin{eqnarray}\label{AdSBH}
M(r_{0})=M-E(r>r_{0})=M-\frac{Q^2}{2r_{0}}.
\end{eqnarray}
With (14) and (16), we arrive at the relation
\begin{eqnarray}\label{AdSBH}
\frac{C(r_{0})}{4\pi M(r_{0})}=\frac{2\pi r_{0}}{4\pi (M-\frac{Q^2}{2r_{0}})}=\frac{1}{\frac{2M}{r_{0}}-\frac{Q^2}{r_{0}^2}}>1.
\end{eqnarray}
It yields the inequality for the horizonless spacetime as
\begin{eqnarray}\label{AdSBH}
C(r_{0})> 4\pi M(r_{0})
\end{eqnarray}
in accordance with Thorne hoop conjecture,
which states that horizons appear when and only
when the mass $M(r_{0})$ and
circumference $C(r_{0})$ satisfy the relation
$C(r_{0})\leqslant 4\pi M(r_{0})$ \cite{hc21}.
We point out that relation (18) holds for
generic parameters.
Here we analytically prove Thorne hoop
conjecture in spatial regular charged fluid sphere spacetimes.

\section{Conclusions}

We analytically examined the validity of Thorne hoop conjecture
in spatial regular charged curve spacetimes.
We took the natural assumption that the matter energy density
is real. On this real energy density assumption, with generic parameters,
we found that horizonless charged fluid sphere should satisfy
the relation (18) in accordance with Thorne hoop conjecture.
In summary, we provided an analytical proof on Thorne hoop
conjecture in the spatially regular static
charged fluid sphere spacetimes.

\begin{acknowledgments}

This work was supported by the Shandong Provincial Natural Science Foundation of China under Grant
No. ZR2018QA008. This work was also supported by a grant from Qufu Normal University
of China under Grant No. xkjjc201906.

\end{acknowledgments}

\end{document}